# THE MAGNETIC COMPENSATION SCHEME OF THE FCC-ee DETECTORS

M. Koratzinos[1], MIT and K. Oide, KEK and CERN

*Abstract*

A crucial part of the design of an FCC-ee detector is the minimisation of the disruption of the beam due to the presence of a large and powerful detector magnet. Indeed, the emittance blow-up of the few meters around the interaction point (IP) at lower energies is comparable to the emittance introduced by the rest of the 100 km ring. Vertical emittance is the single most important factor in achieving high performance (luminosity, in this case) in a modern $e^+e^-$ storage ring such as the FCC-ee. The design adopted is the simplest possible arrangement that can nevertheless deliver high performance: two additional coils per IP side. The performance achieved is such that vertical emittance blow-up will not be a limiting performance factor even in the case of a ring with four experiments, and even in the most demanding energy regime, that of the Z running (about 45 GeV beam energy).

## INTRODUCTION

The FCC project, as a first step, aims to deliver a high-luminosity $e^+e^-$ storage ring in a range of energies from 45 to 182.5 GeV per beam (FCC-ee) [1] [2]. It incorporates a "crab waist" scheme to maximize luminosity at all energies [3] [4]. This necessitates a crossing angle between the electron and positron beams, which is ±15 mrad in the horizontal plane. The detector solenoid envisaged is a large 2 T coil. No magnetic elements can be present in the region approximately ±1.2 m from the interaction point (IP), to leave space for the particle tracking detectors and the luminosity counter.

Therefore, beam electrons experience the full strength of the detector magnetic field close to the IP. The resulting vertical kick needs to be reversed and this is performed in the immediate vicinity. This vertical bump, however, leads to vertical dispersion and an inevitable increase of the vertical emittance of the storage ring, which we here try to minimize. The effect is most important at the Z energies. The vertical emittance budget [1] varies between 1 pm (Z energies) and 2.9 pm (top energies).

Moreover, the very low vertical $\beta^*$ of the machine necessitates that the final focusing quadrupoles have a distance from the IP ($L^*$) of 2.2 m and therefore are inside the main detector solenoid. The final focus quadrupoles should reside in a region with very low residual magnetic field: the equivalent roll of the quad due to the overlapping solenoid should be much smaller than the value assumed for quad misalignment (0.1 mrad), leading to $\int_{2.2}^{3.6} B_z ds \ll 3 \times 10^{-2} Tm$, which corresponds to a vertical emittance blow up of 0.05 pm (The effect increases quadratically with the residual field).

Another obvious requirement is that any magnetic elements should not be in the way of physics sub-detectors, and therefore are made as compact as possible, which in turn increases fringe fields and, therefore, dispersion. In order not to compromise the physics goals of the experiment, the area allowed for magnetic elements is inside a 100 mrad cone at the IP along the median path of electrons and positrons (defined as the Z axis, the direction of the experiment solenoid field).

## THE REQUIREMENTS

We here summarise the list of requirements for the compensation scheme:
1. All elements within a 100 mrad cone
2. Vertical emittance blow-up (cumulative for all IPs) less than 1 pm.
3. The integral $\int B_z ds$ should vanish…
4. and so should the integral $\int B_x ds$, so that any vertical dispersion would not leak to the rest of the ring.
5. $\int B_z ds$ in the vicinity of final focus quads should be much less than $3 \times 10^{-2} \, Tm$.

All the above (conflicting) requirements are satisfied with the design presented here. This work follows from previously presented work in the subject [5] [6], optimised, simplified and improved.

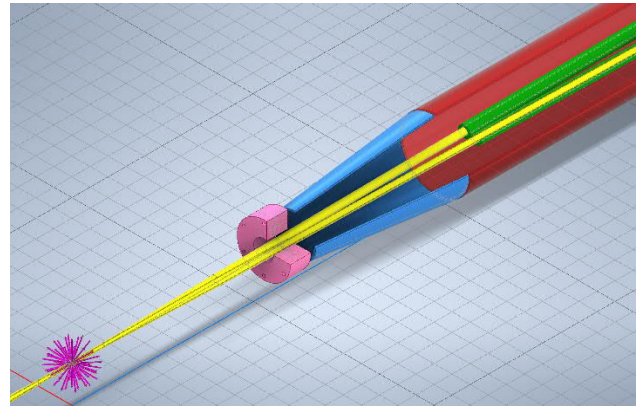

*Figure 1*: The design of the compensation scheme, orthographic three-quarter view. Visible elements are: The IP (magenta), the beam pipe (yellow), the luminosity counter (pink), the compensating solenoid (blue), the screening solenoid (red). The final focus quadrupoles QC1L1 can be seen in green. The detector solenoid has been omitted for clarity.

This analysis is performed for the immediate region around the IP of ±3 m.

---

[1] Currently at CERN

## THE COMPENSATION SCHEME

The beam-stay-clear area in the vicinity of the interaction region is ±12 mm. This allows for a compact *beam pipe* of 30mm in diameter.

Physics reasons dictate the position of the *luminosity counter*: the overall rate of Bhabha events at the Z peak cannot be too much smaller than the Z to hadrons rate. This effectively fixes the position of the front face of the luminometer at a distance of 1076 mm from the IP (the depth of the calorimeter is 116 mm). This forces the first magnetic element of the compensation scheme to start at a distance of 1230 mm from the IP.

The compensation scheme comprises two magnetic elements (solenoids) along the Z axis. The first is a *compensating solenoid* with a negative field compared to the detector solenoid field and the second is a *screening solenoid*, a longer coil that screens the final focus quadrupoles from the detector solenoid field. The presence of two elements makes it possible to minimize the $\int B_z ds$ and $\int B_z ds$ integrals at the same time.

The *detector solenoid* is a cylinder with an inner radius of 376 cm and an outer radius of 382 cm. Its half-length is 400 cm. There is currently no end yoke design, so the field is not as uniform as with an iron return yoke (at 3 m from the IP the field has dropped to about 1.6T from 2T at the IP. This analysis will be updated when the detector magnet design is finalized, but the essence of the analysis and the results presented here will not change.

The *final focus quadrupoles* start at a distance of 2.2m from the IP.

Since large fields are required, the coils mentioned in this work would all be superconducting. A thin, non-load-bearing cryostat is envisaged, as well as a strong load-bearing skeleton and the space is provided for.

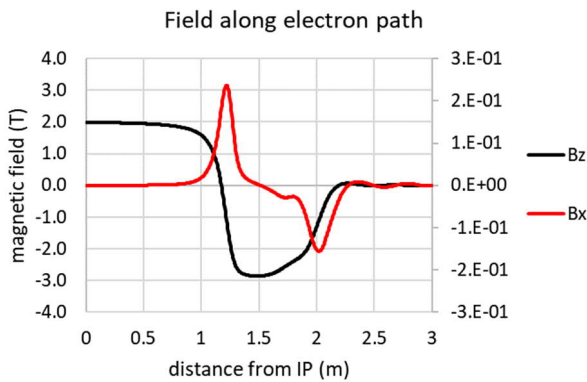

*Figure 2*: *The field profile seen by an electron from the IP up to a distance of 3 m (still inside the detector solenoid). During the first meter or so the electron sees the full detector solenoid field, then the field reverses, thanks to the compensating solenoid, and it finally approaches zero at the tip of the final focus quadrupoles (at 2.2m from the IP). $B_z$ varies between +2T and -2.9T (left scale, black) whereas $B_x$ between 210mT and -160 mT (right scale, red).*

### The screening solenoid

The *screening solenoid* is a thin solenoid producing a field equal and opposite to the detector solenoid, that screens the final focus quadrupoles. It starts at 2000mm from the IP and extends all the way to the endcap region of the detector, at 5.2 m from the IP. Its outer radius is 195mm and has 340 turns. The pitch varies from 7 to 13 mm. The conductor cross-section is 2 by 10 mm and the total current 9980 A, corresponding to a current density of 499 A/mm2. NbTi technology is adequate for this device. Its maximum standalone field is about -1.59T.

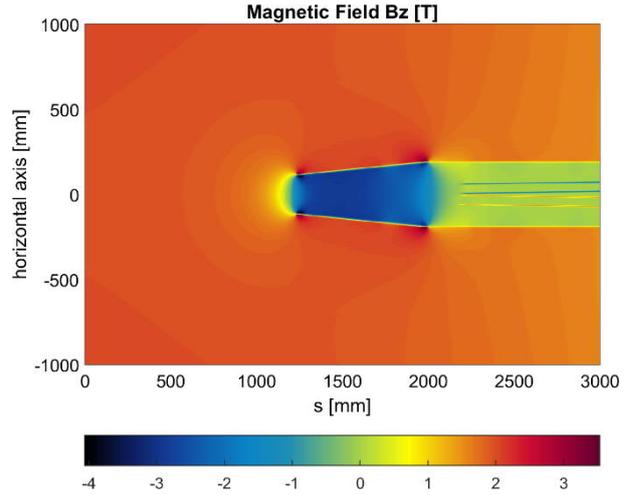

*Figure 3*: *Bird's eye view of the right side of the compensation scheme. The longitudinal component of the magnetic field is shown in the region y=(-1,1m) and z=(0,3m) in the vicinity of the compensating solenoid (blue, -3T), screening solenoid (green, 0T), final focus quadrupoles (just visible in blue and yellow), all in the +2T solenoidal field of the experiment (orange). The IP is at (0,0)*

### The compensating solenoid

The *compensating solenoid* sits in front of the screening solenoid, has a field higher than that of the detector solenoid, so that the magnetic field integral seen by the beam is zero. The length of this solenoid is 77cm, its front face is at 1230mm from the IP, its back face at 2000mm, and its stand-alone strength is -4.77 T. It is tapered: its outer diameter at the front tip is 118 mm and at the back tip 195mm. This leaves space for a thin cryostat of 5 mm depth up to the allowed 100 mrad cone. The coil has 162 turns and the pitch varies between 2.5mm and 6 mm. The conductor is assumed to be 2 by 10 mm and the total current is 19880 A corresponding to 994 Amm$^{-2}$. This current density is beyond the capability of NbTi conductors, so HTS should be used. To be able to use NbTi conductor, the conductor cross-section area should increase by 50%, which is a minor modification.

The different elements of the design can be seen in *Figure 1*. In *Figure 2* the field components in the x (horizontal), and z (longitudinal) direction along the electron path are shown. *Figure 3* shows the map of the longitudinal component of the magnetic field.

## ANALYSIS AND MINIMIZATION

All magnetic design was performed using the *Field* suite of programs [7]. The vertical emittance blow-up was calculated analytically using the equations described in [6] in an excel sheet. This allowed for rapid progress and convergence of the minimization process, which was done to a large extend empirically. The sizes of the coils were given and what was minimized was the pitch of the coils along their length and the current in the conductor.

Only after the best configuration was identified were the exact field maps transferred to the SAD suite of programs, where the exact value of the emittance blow-up was calculated. Some important optics functions can be seen in *Figure 4*.

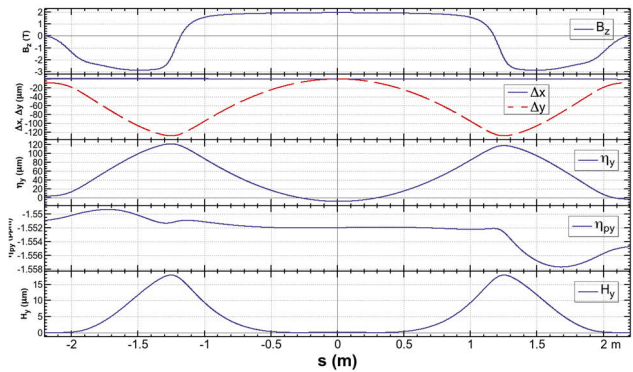

*Figure 4: Optics functions in the area ±2m from the IP. From top to bottom: longitudinal magnetic field, closed orbit deviation from the tilted straight line going through the IP, vertical dispersion, vertical momentum dispersion, $\mathcal{H}_y$ (vertical emittance generation function).*

The emittance blow up of the optimized setup was 0.24 pm at a beam energy of 45.6 GeV. (The simplified excel analysis gave an emittance blow-up of 0.27 pm)

Integral fields are: $\int B_x ds = 2.4 \times 10^{-5}\ Tm$, $\int B_z ds = 5.8 \times 10^{-2}\ Tm$ and in the vicinity of the final focus quadrupole (QC1L1, from 2.2 to 3.6 m from the IP) $\int_{2.2}^{3.6} B_z ds = 6.2 \times 10^{-3}\ Tm$. The relatively large $\int B_z ds$ value is due to the uncertainty in the design of the end yoke of the detector magnet; when this is finalized, the compensation can be tuned to keep this value arbitrarily small.

## VARIATION WITH ENERGY AND DETECTOR MAGNETIC FIELD

The vertical emittance blow-up is a strong function of beam energy, $\Delta \varepsilon_y \propto E_{beam}^{-3}$ therefore going from the Z to W running (45 to 80 GeV) the problem reduces by a factor 5.6. Emittance blow-up is also a strong function of detector solenoid field $\Delta \varepsilon_y \propto B_{detector}^5$ therefore if the detector field is increased from 2T to 3T, the emittance blow-up is a factor of 7.6 larger.

## MISALIGNMENT

The above analysis is with perfect alignment. Out of possible misalignments the most dangerous is a (horizontal) tilt of the detector solenoid with respect to the rest of the system (beam, screening and compensating solenoids- which is relatively easy to align as the beam position monitors and the two solenoids of the are in close proximity). Any horizontal tilt of the detector solenoid will generate a horizontal magnetic field component and a vertical orbit distortion and dispersion over the whole ring.

For a 1mrad tilt of the detector solenoid the corresponding uncorrected distortion is unacceptably large. However, a correction on orbit/dispersion/coupling (no the assumption that we can measure them) using dipoles and skew quadruples on a few sextupoles around the IP, gives an acceptable orbit/dispersion, with the resulting vertical emittance at 0.288 pm (20% larger than the perfectly aligned case).

In the actual machine, the measurement of dispersion and coupling at the IP will be difficult, however we can perform the following: Any tilt of the compensation solenoid with respect to the detector solenoid gives rise to a sizable torque on the compensation solenoid (400Nm per mrad, see *Figure 5*). Equipping the compensating solenoid with strain sensors can ensure alignment to 50 μrad, an operation that can be performed at the beginning of every data-taking period.

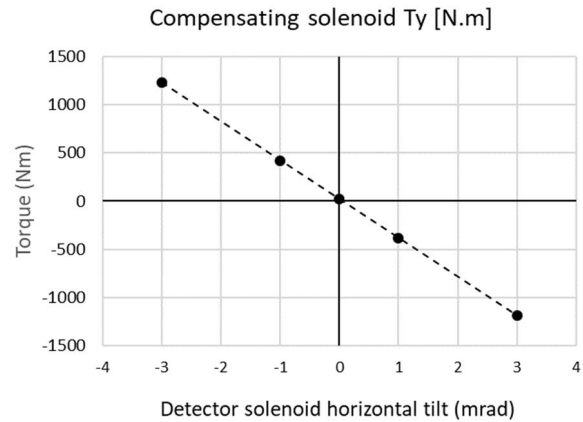

*Figure 5: Torque on the compensation solenoid as a function of relative horizontal tilt to the detector solenoid. The screening solenoid is switched off.*

## CONCLUSIONS

We have designed an efficient and high-performance compensation scheme with two magnetic elements per IP side. We have demonstrated that the very stringent and conflicting requirements are met with this elegant design. The vertical emittance blow up from two IPs is 0.24 pm at the Z energies, compared to the emittance budget of 1 pm. Therefore, the presence of detector solenoids will not impair the performance of the FCC-ee collider, even with existence of 4 IPs.